# Axial-vector form factor of nucleons in the isospin medium from the hard-wall AdS/QCD model.

*Ibrahim Atayev,[1] Shahin Mamedov[2,1]*
[1]*Institute of Physics, Azerbaijan National Academy of Sciences, H. Javid 131, Baku, AZ-1143, Azerbaijan*
[2]*Institute for Physical Problems, Baku State University, Z. Khalilov 23, Baku, AZ-1148, Azerbaijan*

**Abstract**

We study the axial-vector form factor of the nucleons at constant and homogenous isospin chemical potential using holographic QCD. Nucleon mass splitting in such isospin medium is taken into account. According to AdS/CFT correspondence the critical value of the isospin chemical potential $\mu_I$ equals to UV- boundary value of the time component of the bulk gauge field. We calculate the axial-vector form factor of the proton and neutron in such background in the framework of the hard-wall model and plot the form factors for the isospin chemical potential critical value $\mu_I = m_\pi$.

## I. Introduction

In recent years, it has been found the correspondence between the classical gravitational theory in a higher dimensional space-time and conformal field theory on the boundary of this space-time. The Anti de Sitter Field/Conformal Field Theory (AdS/CFT) correspondence [1, 2, 3] allows solving those problems, which cannot be solved within the perturbation theory in the quantum chromodynamics (QCD) at low energies. The QCD models, which were based on direct application of the AdS/CFT correspondence, were called holographic QCD or AdS/QCD models. These are mainly two models, which are called the hard-wall and soft-wall models [4-11], and were applied to the calculation of mass spectra, couplings, and decay constants, form factors, and other quantities [12-20], which can be measured in the experiments. Hard and soft-wall models widely applied for nuclear medium studies as well, which are arise in the collision of heavy ions or protons at high energies. Depending on the energy of colliding particles, this nuclear medium can be in confinement or in the deconfinement (or quark-gluon plasma) phases. In the holographic QCD models, each phase is described by specific gravitational background. In the confined phase of the nuclear matter, the nucleons of the medium are still whole and the quarks and gluons inside them are confined. This phase in the dual gravity theory is described by the thermal AdS space (tAdS) [21, 22]. Dual gravity for the confinement phase containing the quark fields as well was found in [22] and the background geometry was named the thermal charged AdS space (tcAdS). To study analytically (or even numerically) the effects and phenomena in a nuclear matter are highly complicated and different simplifications of the medium are available. The most simplified model for the nuclear medium is the isospin medium, where the temperature and quark number density quantities are turned off. The only quantity describing the medium is the isospin chemical potential $\mu_I$. Such a model enables us to separate and qualitatively study the effects, which are caused by an isospin interaction of the particles with the medium. In the holographic hard-wall model the background geometry for the isospin medium is reduced from the tcAdS space to the ordinary AdS space. Isospin medium was applied for the meson mass splitting due to isospin interaction study in the Refs. [23-26]. Nucleons in the constant isospin background were considered in the Ref. [24, 25] and the chiral symmetry breaking in the isospin field was studied in the Ref. [23]. Splitting of meson decay constants because of interaction with the medium's isospin was determined in the Ref. [25].

Within the hard- and soft-wall models framework in the Refs. [14, 17, 40] it was investigated the nucleon's axial-vector form factor $G_A(Q^2)$ in the vacuum. The temperature dependence of this form factor was studied in the Ref. [39]. Main advantage of this model is that, there is no limitations on the internal of the transferred momentum square $Q^2$ and therefore, the $G_A(Q^2)$ form factor was calculated for all values of this variable, while non-holographic models is applicable only at small or only at large values of $Q^2$. Obviously, the isospin medium affects the $G_A(Q^2)$ form factor as well, since the nucleons have a splitting due to the isospin interaction with the medium. In the result of this interaction this form factor should be different for the neutron and proton states, since these states have different values of the isospin projection. Here we aim to investigate splitting of the $G_A(Q^2)$ form factor in the result of isospin interaction using the hard-wall model.

## ISOSPIN MEDIUM SETUP

Let us briefly describe the isospin medium in the hard-wall model of the AdS/QCD following Refs. [23-26]. As was shown in these works in the isospin medium the background geometry for the dual gravity theory is given by the 5- dimensional (5D) anti-de-Sitter (Ad$S_5$) space-time which in Poincare coordinates has the metric in the following simple form:

$$ds^2 = g_{MN}dx^M dx^N = \frac{1}{z^2}(\eta_{\mu\nu}dx^\mu dx^\nu - dz^2). \qquad (1)$$

The fifth coordinate $z$ extends from 0 to $\infty$, values which are called the ultraviolet (UV) and the infrared (IR) boundaries of the AdS space, respectively. $\eta_{\mu\nu}$ is the metric tensor of the 4D Minkowski space ($\eta_{\mu\nu} = diag(1,-1,-1,-1,)$, $\mu,\nu = 0,1,2,3$). The hard-wall model of AdS/QCD is based on the two cut-offs of the $z$ coordinate first, one of which is the cut-off at the bottom of the AdS space by small $\epsilon$ ($\epsilon \to 0$) in order to avoid singularity of the metric (1) at the $z \to 0$ limit and another one is the cut-off at the top of this space, i.e. at some $z_m$ value of this coordinate, which is considered as the free parameter of the model. The $z_m$ parameter is taken as $z_m = 1/\Lambda_{QCD}$ ($\Lambda_{QCD}$ corresponds to the confinement scale of QCD). The isospin matter can be described by two diagonal subgroup elements of the flavor group and it can be further decomposed into the symmetric or antisymmetric combination

$$V_t^3 = \frac{1}{2}(L_t^3 + R_t^3), \qquad A_t^3 = \frac{1}{2}(L_t^3 - R_t^3).$$

Here $L$ and $R$ are the gauge fields introduced in the $SU_L(2)$ and $SU_R(2)$ groups. where the superscript 3 indicates the third component of the nucleon's isospin and the $t$ subscript shows on time component. In the nuclear medium the boundary value of $V_t^3$ corresponds to the difference of the isospin chemical potentials of proton and neutron. We chose the time components of the vector and axial-vector fields: $A_t^3 = 0, V_t^1 = V_t^2 = 0$, as was made in the [28] in order to get anisotropic isospin directed along the third axis. We adopt $V_t^3$ as critical value of chemical potential $\mu_I$ and we introduce the isospin chemical potential $\mu_I$ as a UV- boundary value of the time component of the gauge field of $SU_V(2)$ symmetry as $V_t^3(z)|_\epsilon = \mu_I$,

When we turn on the isospin chemical potential $\mu_I$ at zero baryon number density the pion condensation is expected to occur at a critical point. Son and Stephanov showed that, using the chiral Lagrangian at $O(p^2)$ the phase transition to the pion condensation phase is the second-order, and critical value of $\mu_I$ is equal to the pion mas $\mu_I = m_\pi$ [27, 28, 34].

## II. NUCLEONS IN THE ISOSPIN MEDIUM.

Nucleons in AdS/QCD are introduced by means of two bulk spinor $\psi_1$ and $\psi_2$, which are necessary for the description of the left and right components of this particles [31,32]. Action for the 5D Dirac field in the 5 D-dimensional AdS space has a form:

$$S = \int d^5x \sqrt{g}\, [i\bar{\psi}_1 e_A^M \Gamma^M D_M \psi_1 - m_1 \bar{\psi}_1 \psi_1 + m_2 \bar{\psi}_2 \psi_2, m_1 \leftrightarrow -m_2)], \qquad (2)$$

where $g$ denotes $g = |det g_{MN}|(M, N = 0,1,2,3,5)$. For the AdS space (1) an inverse vielbein is $e_A^M = z\delta_A^M$. The covariant derivative for this field is $D_M = \partial_M + \frac{1}{8}\omega_{MAB}[\Gamma^A, \Gamma^B] - iV_M$. Non-zero components of the spin connection are $\omega_\mu^{5A} = -\omega_\mu^{A5} = \frac{1}{z}\delta_\mu^A$. The 5D-dimensional $\Gamma^A$ matrices are defined as $\Gamma^A = (\gamma^\mu, -i\gamma^5)$ and obey anti-commutation relations $\{\Gamma^A, \Gamma^B\} = 2\eta^{AB}$. For consistency with the chirality of dual boundary operator the sign of the mass is chosen positive ($m_1 > 0$) for the $\psi_1$ and negative ($m_1 < 0$) for the $\psi_1$. This chirality is related to the chirality of the boundary fermion, i.e. proton and neutron [20, 21]. Following the AdS/CFT relation

$$(m_{1,2})^2 = (\Delta - 2)^2,$$

one should take $m_1 = \frac{5}{2}$ and at the same time $m_2 = -\frac{5}{2}$. Dual composite fermionic operators $\psi_{1,2}$ have conformal dimension $\Delta = \frac{9}{2}$.

In the description of the isospin medium the quarks and nucleons are differrent from the ordinary quarks and nucleons in the vacuum. In order to distinguish such particles from the ordinary ones we use the terminology, iso-particle which was introduced in [26]. Since the proton and neutron are different isospin states of the nucleons, they have non-zero isospin. The isospin medium in this model is introduced as a constant isospin background and the bayron number density for this medium were turned off. Such background was constructed in [23,24] as a one created by isoquarks which has only different interaction energy with isospin background and in the result the masses of the proton and neutron are different.

From the action (3) one obtains the equation of motion for the bulk Dirac fields:
$$[ie_A^M \Gamma^A D_M - m_{1,2}]\psi_{1,2} = 0. \tag{4}$$

Introducing the Fourer mode of the fermion:
$$\psi_{L,R}^{1,2}(x,z) = \int \psi_{L,R}^{1,2}(p) f_{R,L}^{1,2}(p,z) e^{-ipx} d^4p \tag{5}$$

and the Weyl spinor representation with $\psi_L^{1,2} = \gamma^5 \psi_L^{1,2}$ and $\psi_R^{1,2} = -\gamma^5 \psi_R^{1,2}$

$$\psi^{1,2}(p) = \begin{pmatrix} \psi_L^{1,2}(p) \\ \psi_R^{1,2}(p) \end{pmatrix}. \tag{6}$$

The boundary fields $\psi_{L,R}^{(1,2)}(p)$ are related to each other by the 4-dimensional Dirac equation:
$$\gamma^\mu p_\mu \psi_{L,R}^{1,2}(p) = |p|\psi_{R,L}^{1,2}(p). \tag{7}$$

Where the subscripts $L$ and $R$ denote the 4-dimensional chirality. If one takes the normalizable modes as $f_L^1$ and $f_R^1$ for $\psi_L^1$ and $\psi_R^1$ respectively, the chirality of the $SU(2)_L \times SU(2)_R$ flavor group can be associated with the 4-dimensional chirality. In terms of the Weyl fermions, the equations in (4) are reduced to the following ones:

$$\begin{pmatrix} \partial_z - \frac{\Delta}{z} & 0 \\ 0 & \partial_z - \frac{4-\Delta}{z} \end{pmatrix} \begin{pmatrix} f_L^1 \\ f_L^2 \end{pmatrix} = -(|p| - V_t) \begin{pmatrix} f_R^1 \\ f_R^2 \end{pmatrix} \tag{8}$$

$$\begin{pmatrix} \partial_z - \frac{\Delta}{z} & 0 \\ 0 & \partial_z - \frac{4-\Delta}{z} \end{pmatrix} \begin{pmatrix} f_R^2 \\ f_R^1 \end{pmatrix} = (|p| - V_t) \begin{pmatrix} f_L^2 \\ f_L^1 \end{pmatrix} \tag{9}$$

These matrix equation can be further reduces to the symmetric and anti-symmetric combinations, which describe the parity-even and parity-odd excitations under $1 \leftrightarrow 2$ and simultaneously $L \leftrightarrow R$ change. As a result, the lowest nucleon spectra corresponding to the proton and neutron, which are parity-even states, are described by the lowest excitation of the symmetric combination $f_L^1 + f_R^2$ together with $f_R^1 - f_L^2$. On the other hand, the parity-odd states are represented as $f_L^1 - f_R^2$ and $f_R^1 + f_L^2$. In order to investigate the parity-even mass spectra, one can impose $f_L^1 = f_R^2$ and $f_R^1 = -f_L^2$, then the above two matrix equations, (8) and (9), reduce to the same matrix equation

$$\begin{pmatrix} \partial_z - \frac{\Delta}{z} & 0 \\ 0 & \partial_z - \frac{4-\Delta}{z} \end{pmatrix} \begin{pmatrix} f_L^1 \\ f_R^1 \end{pmatrix} = \begin{pmatrix} -(|p| - V_t) & 0 \\ 0 & |p| - V_t \end{pmatrix} \begin{pmatrix} f_R^1 \\ f_L^1 \end{pmatrix}. \tag{10}$$

Furthermore, since the nucleon has an isospin charge, it can interact with the background isospin matter. If the $n$–th excited mode of $f_{L,R}^1$ todenote by $f_{L,R}^{1(n,\pm,\pm)}$ where the first and second sign imply the parity and isospin quantum number respectively, the parity-even state satisfying (9) can be further decomposed, depending on the isospin charge, into

$$\begin{pmatrix} \partial_z - \frac{\Delta}{z} & 0 \\ 0 & \partial_z - \frac{4-\Delta}{z} \end{pmatrix} \begin{pmatrix} f_L^{1(n,+,+)} \\ f_R^{1(n,+,+)} \end{pmatrix} = \begin{pmatrix} -(|p| - \frac{V_t^3}{2}) & 0 \\ 0 & |p| - \frac{V_t^3}{2} \end{pmatrix} \begin{pmatrix} f_R^{1(n,+,+)} \\ f_L^{1(n,+,+)} \end{pmatrix}, \tag{11}$$

$$\begin{pmatrix} \partial_z - \frac{\Delta}{z} & 0 \\ 0 & \partial_z - \frac{4-\Delta}{z} \end{pmatrix} \begin{pmatrix} f_L^{1(n,+,-)} \\ f_R^{1(n,+,-)} \end{pmatrix} = \begin{pmatrix} -(|p| + \frac{V_t^3}{2}) & 0 \\ 0 & |p| + \frac{V_t^3}{2} \end{pmatrix} \begin{pmatrix} f_R^{1(n,+,-)} \\ f_L^{1(n,+,-)} \end{pmatrix}. \tag{12}$$

On the other hand, the parity-odd states are governed by following equations: As the result is obtained from equations (11) and (12).

$$\begin{pmatrix} \partial_z - \frac{\Delta}{z} & 0 \\ 0 & \partial_z - \frac{4-\Delta}{z} \end{pmatrix} \begin{pmatrix} f_L^{1(n,-,+)} \\ f_R^{1(n,-,+)} \end{pmatrix} = \begin{pmatrix} -(|p| - \frac{V_t^3}{2}) & 0 \\ 0 & |p| - \frac{V_t^3}{2} \end{pmatrix} \begin{pmatrix} f_R^{1(n,-,+)} \\ f_L^{1(n,-,+)} \end{pmatrix}, \tag{13}$$

The lowest excitation modes, when $n = 1$, the proton and neutron, can be represented by the $f_{L,R}^{1(1,+,+)}$ and $f_{L,R}^{1(1,+,-)}$ profile functions. Since the lowest excitations have only the parity-even states, there is no the $f_{L,R}^{1(1,-,\pm)}$ profiles. For the higher resonances, they can have even and odd parity states.

From the above matrix equation, one can easily derive two second order differential equations for $f_L^{1(n,\pm,\pm)}$ isospin states:

$$(\partial_z^2 - \frac{4}{z}\partial_z + \frac{(5-\Delta)\Delta}{z^2})f_L^{1(1,\pm,+)} = -(|p| - \frac{V_t^3}{2})^2 f_L^{1(1,\pm,+)} \text{ for proton}, \tag{15}$$

$$(\partial_z^2 - \frac{4}{z}\partial_z + \frac{(5-\Delta)\Delta}{z^2})f_L^{1(1,\pm,+)} = -(|p| + \frac{V_t^3}{2})^2 f_L^{1(1,\pm,-)} \text{ for neutron}. \tag{16}$$

Similarly, the differential equations for the $f_R^{1(n,\pm,\pm)}$ profiles are the following ones:

$$(\partial_z^2 - \frac{4}{z}\partial_z + \frac{(5-\Delta)\Delta}{z^2})f_R^{1(1,\pm,+)} = -(|p| - \frac{V_t^3}{2})^2 f_R^{1(1,\pm,+)} \quad \text{for proton}, \tag{17}$$

$$(\partial_z^2 - \frac{4}{z}\partial_z + \frac{(5-\Delta)\Delta}{z^2})f_R^{1(1,\pm,+)} = -(|p| + \frac{V_t^3}{2})^2 f_R^{1(1,\pm,-)} \quad \text{for neutron}. \tag{18}$$

The mass of nucleon is given by $p$ and is find from the following two boundary conditions on profile functions:

$$f_L^{1(1,\pm,\pm)}(0) = 0 \text{ and } f_R^{1(1,\pm,\pm)}(z_{IR}) = 0. \tag{20}$$

Now, let us take into account the mass splitting of nucleons in the isospin medium. The nucleon mass is given by $p_0$ at $V_t^3 = 0$ is the same for the proton and neutron. In the isospin medium with nonzero $V_t^3$, the proton and neutron masses are shifted due to the isospin interaction. Since the isospin medium has a constant $V_t^3$, the boundary conditions (20) lead to $p_0 = p_p - \frac{V_t^3}{2}$ for proton and $p_0 = p_n + \frac{V_t^3}{2}$ for neutron. In order to study the form factor splitting in the isospin medium we should fix value of $V_t^3$. As is known [27, 28, 34], at the isospin chemical potential value $\mu_I = m_\pi$ it occurs pion condensation. So, taking $V_t^3 = m_\pi$, we can study the form factor splitting in the pion condensation state. Proton and neutron masses in this isospin medium are given by:

$$p_P = p_0 + \frac{m_\pi}{2}, \qquad p_n = p_0 - \frac{m_\pi}{2}. \tag{22}$$

Solutions of the (15)-(18) equations obeying the UV boundary conditions are the Bessel functions of first kind:

$$f_{L,R} = C_{1,2} z^{\frac{5}{2}} J_{M \mp \frac{1}{2}}(|p_N|, z), \tag{23}$$

where $C_{1,2}$ are normalization constants and $p_N$ is the $p_{P,n}$. The value of $m_5$ can be found from the relation $M = \Delta_{\frac{1}{2}} - 2$, where scaling dimension $\Delta_{\frac{1}{2}}$ for the composite baryon operator is $\Delta_{\frac{1}{2}} = \frac{9}{2}$ [18] and $|M| = \frac{5}{2}$. Consequently, for the $\psi^{1,2}$ spinors the $M$ mass have the values $M = \pm \frac{5}{2}$ correspondingly. Thus, the $f_{L,R}^1$ and $f_{L,R}^2$ profile functions are given by

$$f_L^1 = C_1 z^{\frac{5}{2}} J_2(|p_N|, z) \quad f_R^1 = C_2 z^{\frac{5}{2}} J_3(|p_N|, z),$$
$$f_L^2 = C_2 z^{\frac{5}{2}} J_3(|p_N|, z) \qquad f_R^2 = C_1 z^{\frac{5}{2}} J_2(|p_N|, z). \tag{24}$$

As seen as from (24) $f_{L,R}^1$ and $f_{L,R}^2$ are related one with another:

$$f_L^1 = f_R^2, \qquad f_R^1 = -f_L^2. \tag{25}$$

The normalization constants $C_{1,2}$ in (24) are equal for the $n$-th excited state were found in [17]:

$$C_{1,2}^n = C^n = \frac{\sqrt{2}}{z_m J_2(M_n z_m)} \tag{26}$$

$M_n$ is the Kaluza-Klein mass spectrum of exited states and is expressed in terms of zeros $\alpha_n^{(3)}$ of the Bessel function $J_3$:

$$M_n = \frac{\alpha_n^{(3)}}{z_m}.$$

### Axial-Vector Current in QCD

Isovector axial-vector current of nucleons is defined in the following [35]:

$$j^{\mu,a}(x) = \bar{\psi}(x) \gamma^\mu \gamma^5 \frac{\tau^a}{2} \psi(x). \tag{27}$$

Here $\psi(x)$ denotes the doublet of $u$ and $d$ quarks $\psi = \binom{u}{d}$ and $\tau^a$ are the Pauli matrices describing isospin. A matrix element of the isovector current (27) between one-nucleon states is defined is defined in terms of two form factors [35]:

$$\langle N(p')|j^{\mu,a}(0)|N(p)\rangle = \bar{u}(p') \left[\gamma^\mu \gamma^5 G_A(q^2) + \frac{q^2}{2m_N} \gamma^5 G_p(q^2)\right] \frac{\tau^a}{2} u(p) \tag{28}$$

Here $m_N$ is nucleon mass, $q_\mu = p' - p_\mu$ total momentum in the interaction vertex. $G_A^2(q^2)$ and $G_p(q^2)$ are called the axial-vector and pseudoscalar form factors respectively.

### Axial-vector field in the AdS Space

In addition to (3) we have the kinetic term of the axial-vector field in the action:

$$S_A = \int d^5 x \sqrt{g} Tr\left(-\frac{F_A^2}{2g_5^2}\right). \tag{29}$$

Where $F_{MN}^A = \partial_M A_N - \partial_N A_M$, $g_5^2 = \frac{12\pi^2}{N_c}$. The transverse part of the axial-vector field can be written as $A_\mu(q,z) = A(q,z) A_\mu^0(q)$. Near the UV boundary an equation of motion for this field coincide with one for a vector field and in the $A_z = 0$ gauge it has a form [16-18]:

$$z\partial_z \left(\frac{1}{z}\partial_z A(q,z)\right) + q^2 A(q,z) = 0. \tag{30}$$

UV and IR boundary conditions on this solution are $A(q, z = \epsilon) = 1$ at $\epsilon \to 0$ and $\partial_z A(q, z = z_m) = 0$ correspondigly. Solution of the equation (30) is expressed via the first and second kind Bessel functions $J_m$ and $Y_m$ [14]:

$$A(q,z) = \frac{\pi}{2}\left[\frac{Y_0(qz_m)}{J_0(qz_m)} J_1(qz) - Y_1(qz)\right].$$

### The Chiral Symmetry Breaking by Scalar Field

Action for the scalar $X$ field is the usual one for the scalar field in the five-dimesional background geometry (1):

$$S = -\int d^5 x \sqrt{g} Tr[|DX|^2 + 3|X|^2]. \tag{31}$$

The covariant derivative $D_M$ includes the interaction of this field with the $A_L$ and $A_R$ gauge fields

$$D_M = \partial_M X - i(A_L)_M X + iX(A_R)_M.$$

Its interaction with the spinor fields will be written in separate terms in the interaction Lagrangian. The $X$ field is written in the form: $X = v(z)\exp[i\sqrt{2}\pi^a T^a]$, where the $\pi^a$ field in the dual QCD describe pions. In the free field limit the solution of the equation of motion obtained from the action (10) for $v(z)$ has a form [17]:

$$v(z) = \frac{1}{2}\frac{\sqrt{N_C}}{2\pi} m_q z + \frac{1}{2}\frac{2\pi}{\sqrt{N_C}} \sigma z^3, \tag{32}$$

where $m_q$ is the mass of bare light quarks and $\sigma$ is the value of the quark condensate and $N_C = 3$.

### Axial-vector isovector form factor of nucleons

The $G_A$ form factor in the AdS/QCD models framework will be extracted from the 5D action integral for the interaction between the axial-vector, scalar, and fermion fields in the bulk of AdS space[14]:

$$S_{int}^A = \int d^5 x \sqrt{g} \mathcal{L}_{int}^A(x,z). \tag{33}$$

Here the interaction Lagrangian $\mathcal{L}_{int}^A(x,z)$ describes several kind of interactions between $A_M, X$ and $\psi_{1,2}$ fields in the bulk and consists of corresponding terms in the AdS space [14,15,16,17,18]. Let us list the bulk interactions, that contribute to the $G_A(q^2)$ form factor:
  a. Minimal coupling term:

$$L = \bar{\psi}_1 \Gamma^M (A_L)_M \psi_1 - \bar{\psi}_2 \Gamma^M (A_R)_M \psi_2 = \frac{1}{2}(\bar{\psi}_1 \Gamma^M A_M \psi_1 - \bar{\psi}_2 \Gamma^M A_M \psi_2). \tag{34}$$

b. Magnetic gauge coupling term:
$$L = ik_1\{\bar{\psi}_1 \Gamma^{MN}(F_L)_{MN}\psi_1 - \bar{\psi}_2 \Gamma^{MN}(F_R)_{MN}\psi_2\} =$$
$$= k_1\{\bar{\psi}_1 \Gamma^{MN} F_{MN} \psi_1 + \bar{\psi}_2 \Gamma^{MN} F_{MN} \psi_2\}. \tag{35}$$

c. Triple interaction term, which was introduced in [14]:
$$L = \frac{g_Y}{2}[\bar{\psi}_1 X \Gamma^M (A_L)_M \psi_2 - \bar{\psi}_2 X^+ \Gamma^M (A_R)_M \psi_1 + \text{h.c}] =$$
$$= g_Y[\bar{\psi}_1 X \Gamma^M A_M \psi_2 + \bar{\psi}_2 X^+ \Gamma^M A_M \psi_1]. \tag{36}$$

Having an interaction Lagrangian in the bulk we can derive a holographic expression for the $G_A$ form factor. The AdS/CFT correspondence in our case matches the axial-vector current of the bulk fermions with the axial-vector current of nucleons in the boundary QCD and at the same time it relates the bulk axial-vector field with the axial-vector meson in this boundary theory. According to AdS/CFT correspondence, the generating functional $Z_{AdS}$ which is defined as an exponent of the classical bulk action

$$Z_{AdS} = e^{iS_{int}}, \tag{37}$$

is equal to generating function $Z_{QCD}$ of the boundary QCD:

$$Z_{AdS} = Z_{QCD}. \tag{38}$$

The above statement allows us to calculate the vacuum expectation value of the nucleons axial-vector current in the boundary QCD by taking variation from the gravity functional $Z_{AdS}$:

$$<J_\mu^a>^{QCD} = -i\frac{\delta Z_{AdS}}{\delta A_\mu^a}|_{A_\mu^a=0}. \tag{39}$$

Formula (39) produces the axial-vector current $J^{5\mu}(p',p) = G_A(q^2)\bar{u}(p')\gamma^5\gamma_\mu\frac{\tau^a}{2}u(p)$, where $G_A(q^2)$ denotes the integral over the z coordinate and is accepted as the axial-vector form factor of nucleons due to the holographic correspondence.

Using (34),(35),(36) we can calculate $S_{int}^i$ in the momentum space. This gives us the following results:

$$S^{(a)} = \frac{1}{2}\int d^4x \int_0^{z_m} dz \sqrt{g}\{\bar{\psi}_1 \Gamma^\mu A_\mu \psi_1 - \bar{\psi}_2 \Gamma^\mu A_\mu \psi_2\} =$$
$$= \frac{1}{2}\int d^4p' d^4p J^{5\mu}(p',p)A_\mu^a(q) \int_0^{z_m} dz \frac{1}{z^4} A(q,z)[F_{1R}^2(p_{P,n},z) - F_{1L}^2(p_{P,n},z)]. \tag{40}$$

1) $S^{(b)} = \frac{1}{4}k_1\int d^4x \int_0^{z_m} dz \sqrt{g}\{\bar{\psi}_1[\Gamma^5,\Gamma_\mu]\partial_5 A_\mu \psi_1 + \bar{\psi}_2[\Gamma^5,\Gamma_\mu]\partial_5 A_\mu \psi_2\} =$
$$= \frac{k_1}{2}\int d^4p d^4p' J^{5\mu}(p',p)A_\mu^a(q) \int_0^{z_m} dz \frac{1}{z^3}\partial_z A(q,z)[F_{1R}^2(p_{P,n},z) + F_{1L}^2(p_{P,n},z)]. \tag{41}$$

2) $S^{(c)} = g_Y \int d^4x \int_0^{z_m} dz \sqrt{g}\{\bar{\psi}_1 X \Gamma^\mu A_\mu \psi_2 + \bar{\psi}_2 X^+ \Gamma^\mu A_\mu \psi_1\} =$
$$= 2g_Y \int d^4p d^4p' J^{5\mu}(p',p)A_\mu^a(q) \int_0^{z_m} dz \frac{1}{z^4} A(q,z) 2v(z) F_{1L}(p_{P,n},z) F_{1R}(p_{P,n},z). \tag{42}$$

According to the holographic formula (39) the total action $S = S^{(a)} + S^{(b)} + S^{(e)}$ will produce the axial-vector form factor $G_A(q^2)$. Taking derivatives over $A_\mu^a(q)$ from the $S^{(i)}$ action terms we shall get contributions $G_A^i(q^2)$ of these terms into the axial-vector form factor $G_A(q^2)$:

$$G_A^{(a)}(q^2) = \frac{1}{2}\int_0^{z_m} dz \frac{1}{z^4} A(q,z)[(f_R^1)^2(p_{P,n},z) - (f_L^1)^2(p_{P,n},z)], \tag{43}$$

$$G_A^{(b)}(q^2) = \frac{k_1}{2}\int_0^{z_m} dz \frac{1}{z^3}\partial_z A(q,z)[(f_R^1)^2(p_{P,n},z) + (f_L^1)^2(p_{P,n},z)], \qquad (44)$$

$$G_A^{(c)}(q^2) = 2g_Y \int_0^{z_m} dz \frac{1}{z^4} A(q,z)v(z)f_L^1(p_{P,n},z)f_R^1(p_{P,n},z). \qquad (45)$$

So, making a numerical integration, we can plot the $G_A(q^2)$ form factor, which is defined as the sum of $G_A^i(q^2)$ terms: $G_A = G_A^{(a)} + G_A^{(b)} + G_A^{(c)}$, for the $Q^2 = -q^2$ domain. In terms of $Q$ dependence the profile function $A(q,z)$ for axial-vector field gets the form below:

$$A(Q,z) = \frac{\pi}{2}Qz\left[\frac{K_0(Qz_m)}{J_0(Qz_m)}J_1(Qz) + K_1(Qz)\right]. \qquad (46)$$

### Numerical analysis

In order to perform the numerical integration of the $G_A(Q^2)$ form factor, the value of light quark mass $m_q$ and the value of quark condensate $\sigma$ were taken $m_q = 0.00234\ GeV$ and $(\sigma)^{\frac{1}{3}} = 0.311\ GeV$ correspondingly [32, 33]. The constant $k_1 = -0.98$ was taken from [32]. The value $g_Y = 9.182$ was taken from [33], which was found at establishing the correct nucleon mass within the hard-wall model having fixed parameters $m_q = 0.00234\ GeV$, $(\sigma)^{\frac{1}{3}} = 0.311\ GeV$ and $z_m = (0.330\ GeV)^{-1}$. We use the (32) expression for $v(z)$, which is included into the $G_A(Q^2)$ form factor expressions (43), (44), (45). Nucleons were taken in the ground state.

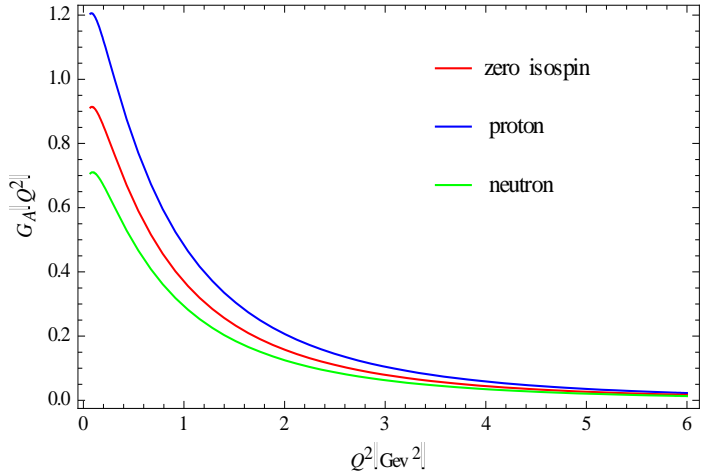

Fig. 1

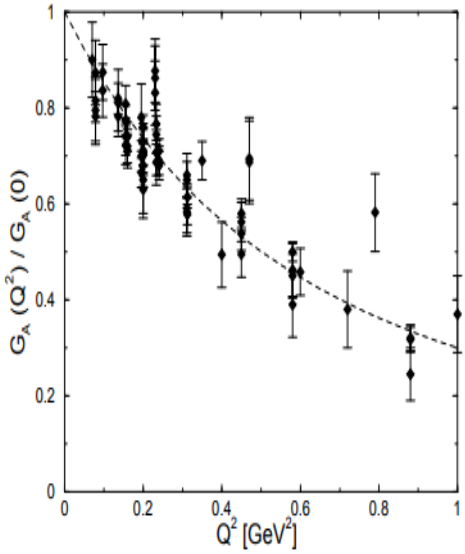 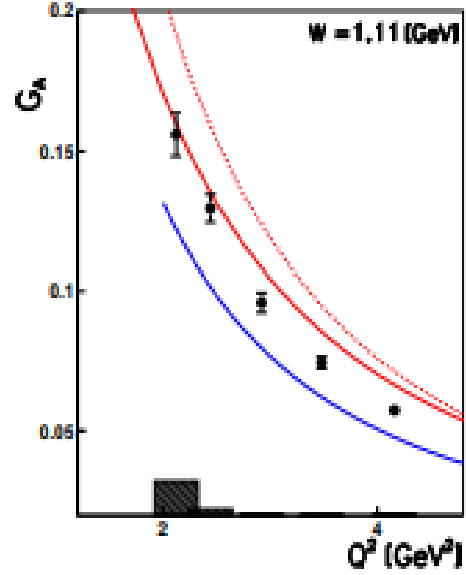

Fig.2                     Fig.3

**Summary**

In this work we studied the splitting of axial-vector form factor of the nucleons in the constant isospin background. Value of the background isospin was chosen the value, at which the pion condensation occur. We plot the $G_A(Q^2)$ form factor in this background for the proton and neutron. Axial-vector form factor for the proton is shifted up, while this formfactor for the neutron is shifted down from the $G_A(Q^2)$ dependence at zero isospin case. The splitting of the axial-vector form factor dependencies occurs due to nucleon mass splitting in the isospin background. Comparison of our result for zero isospin case with the experimental data described in Fig. 2 and Fig. 3 [36, 37], shows that our results are close to the data.

**References**

[1] J. M. Maldacena, Adv. Theor. Math. Math. Phys. **2,** 231 (1998) [Int. J. Theor. Phys. **38,** 1113 (1999)] [arXiv:hep-th/9711200].

[2] E. Witten, Adv. Theor. Math. Math. Phys.**2** ,253 (1998) [arXiv:hep-th/ 9802150 ].

[3] S. S. Gubser, I. R. Klebanov and A.M. Polyakov, Phys. Lett. B **428**,105 (1998) [arXiv:hep-th/9802109].

[4] J. Erlich, E. Katz, D. T. Son and A. M. Polyakov, Phys. Rev. Lett.**95**,261602 (2005) [arXiv: 0501128[hep-ph]].

[5] L. Da Rold and A. Pomarol, Nucl. Phys. B **721**, 79 (2005)[arXiv: 0501218 [hep-ph]].

[6] J. Polchinski and M. J. Strassler, Phys. Rev. Lett. **88,** 031601 (2002) [arXiv:hep-th/0109174].

[7] J. Polchinski and M. J. Strassler, JHEP **0305**, 012 (2003) [arXiv:hep-th/0209211].

[8] S. J. Brodsky and G. F. de Teramond, Phys. Lett. B **582,** 211(2004)[arXiv:hep-th/0310227].

[9] G. F. de Teramond and S. J. Brodsky, Phys. Rev. Lett. **94,** 201601 (2005) [arXiv:hep-th/0501022].

[10] S. J. Brodsky and G. F. de Teramond, Phys. Rev. Lett. **96,** 201601(2006) [arXiv:hep-ph/0602252].

[11] S.J. Brodsky and G. F. de Teramond, Phys. Rev. D **77**, 056007(2008) [arXiv:hep-ph/0707.3859].

[12] B.-H. Lee, C. Park, S. Nam, JHEP 1505, 011 (2015) [arXiv:1412.3097 [hep-ph]]

[13] B.-H. Lee, C. Park, Phys. Lett. B 746, 202 (2015) [arXiv:1503.03615 [hep-th]]

[14] Sh. Mamedov, B. B. Sirvanli, I. Atayev, N. Huseynova, Int. J. Theor. Phys. 56 (2017) 6, 1861 [arxiv:1609.00167 [hep-th]]; I. Atayev, Sh. Mamedov, Russian Physics Journal 64, 12 (2021)

[15] G. Ramalho, Phys.Rev. D 2018) 7, 073002 [arxiv:1707.07206 [hep-ph]]

[16] G. Ramalho and D. Melnikov, Phys. Rev. D 97, 034037 (2018)

[17] T. Gutsche, V. E. Lyubovitskij, I. Schmidt, A. Vega, Phys. Rev. D **77**, 056007 (2012) [arXiv:hep-ph/1204.6612].

[18] Z. Abidin, C. Carlson, Phys. Rev. D, **79** (2009) [arXiv:hep-ph/115003].

[19] H. R. Grigoryan and A. V. Radyushkin, Phys. Lett. B **650** , 421(2007) 0703069.

[20] H. R. Grigoryan and A. V. Radyushkin, Phys. Rev. D **76,** 115007 (2007) [arXiv:hep-ph/0709.0500].

[21] C.P. Herzog, Phys. Rev. Lett. 98, 091601 (2007) [arXiv:hep-th/0608151]

[22] C.A. Ballon Bayona, H. Boschi-Filho, N.R.F. Braga, L.A. Pando Zayas, Phys. Rev. D 77, 046002 (2008) [arXiv:0705.1529 [hep-th]]

[23] B.-H. Lee, C. Park, S.-J. Sin, JHEP 0907, 087 (2009) [arXiv:0905.2800 [hep-th]]

[24] B.-H. Lee, Sh. Mamedov, S. Nam, C. Park, JHEP 1308, 045 (2013)

[25] Sh. Mamedov, Eur. Phys. J. C 76 (2016) 2, 83

[26] B.-H. Lee, Sh. Mamedov, C. Park, Int. J. Mod. Phys. A 29, 1450170 (2014) [arXiv: 1402.6061 [hep-th]]

[27] D. T. Son and M. A. Stephanov, Phys. Atom. Nucl. 64, 834 (2001) [Yad. Fiz. 64, 899 (2001)]

[28] H. Nishihara and M. Harada, Phys. Rev. D 89, 076001 (2014) [arXiv:1401.2928 [hep-ph]]

[29] M.A. Martin Contreras, E.F. Capossoli, D. Li, A. Vega and H. Boschi-Filho, Phys. Lett. B 822 (2021) 136638 [arXiv:2108.05427 [hep-ph]]

[30] A. Cherman, T. Cohen, E. Werbos, Phys. Rev. **C** 79:045203, (2009) arXiv:0804.1096 [hep-ph]

[31] D.K. Hong, T. Inami and H.-U. Yee, Phys. Lett. B 646, 163 (2007) [arXiv:0609270 [hep-ph]]

[32] H. C. Ahn, D. K. Hong, C. Park, S. Siwach, Phys. Rev. D **80,** 054001 (2009) [arXiv: hep-ph/ 0904.3731].

[33] N. Maru, M. Tachibana, Eur. Phys. J. C **63,** 123(2009) [arXiv: hep-ph/0904.3816]

[34] M. Lv, D. Li and S. He, JHEP 11, 026 (2019) [arXiv:1811.03828 [hep-ph]]; X. Cao, H. Liu, D. Li, Phys. Rev. D **102**, 126014 (2020) [arXiv: 2009.00289 [hep-ph]]

[35] G. Ramalho and K. Tsushima, Phys. Rev. D 94, 014001 (2016) [arXiv:1512.01167 [hep-ph]].

[36] V. Bernard, L. Elouadrhiri, U. - G. Meissner, J. Phys. **G28,** R1(2002) [arxiv:0107088[hep-ph]]

[37] Park, K., et al.: CLAS Collaboration. Phys. Rev. C **85** 035208 (2012). [arXiv:1201.0903 [nucl-ex]]

[38] C. Chen, C. S. Fischer, C. D. Roberts, J. Segovia, Phys. Lett. **B 815**, 136150 [arXiv:2011.14026 [hep-ph]]; C. Chen, C. S. Fischer, C. D. Roberts, J. Segovia, "Nucleon axial-vector and pseudoscalar form factors, and PCAC relations" [arXiv:2103.02054 [hep-ph]]

[39] Shahin Mamedov, Narmin Nasibova, "Axial-vector form factor of nucleons at finite temperature from the AdS/QCD soft-wall model" [arXiv:2201.03324 [hep-ph]]

[40] I. Atayev and Sh. Mamedov, Izv. Vuzov. Fizika, v 64, n 12, 88 (2021)